\documentclass{sig-alternate}
\usepackage{mathptmx} 

\usepackage{fancyhdr}
\usepackage[normalem]{ulem}
\usepackage[hyphens]{url}
\usepackage[sort,nocompress]{cite}
\usepackage[final]{microtype}
\usepackage[keeplastbox]{flushend}

\usepackage{amsmath,amssymb,amsfonts}
\usepackage{algorithmic}
\usepackage{graphicx}
\usepackage{textcomp}
\usepackage{xcolor}
\usepackage[separate-uncertainty=true,binary-units=true]{siunitx} 
\usepackage{todonotes}
\usepackage{listings}
\usepackage[frozencache,cachedir=.]{minted} 
\usepackage{braket}

\usepackage{pgfplots}
\pgfplotsset{compat=newest}
\usepgfplotslibrary{units}

\pgfplotsset{colormap={jet}{[1pt]
		rgb(0pt)=(0.0, 0.0, 0.5);
		rgb(1pt)=(0.0, 0.0, 0.535650623885918);
		rgb(2pt)=(0.0, 0.0, 0.589126559714795);
		rgb(3pt)=(0.0, 0.0, 0.62477718360071299);
		rgb(4pt)=(0.0, 0.0, 0.67825311942958999);
		rgb(5pt)=(0.0, 0.0, 0.71390374331550799);
		rgb(6pt)=(0.0, 0.0, 0.76737967914438499);
		rgb(7pt)=(0.0, 0.0, 0.80303030303030298);
		rgb(8pt)=(0.0, 0.0, 0.85650623885917998);
		rgb(9pt)=(0.0, 0.0, 0.90998217468805698);
		rgb(10pt)=(0.0, 0.0, 0.94563279857397498);
		rgb(11pt)=(0.0, 0.0, 0.99910873440285197);
		rgb(12pt)=(0.0, 0.0, 1.0);
		rgb(13pt)=(0.0, 0.0176470588235293, 1.0);
		rgb(14pt)=(0.0, 0.049019607843137254, 1.0);
		rgb(15pt)=(0.0, 0.096078431372549025, 1.0);
		rgb(16pt)=(0.0, 0.12745098039215685, 1.0);
		rgb(17pt)=(0.0, 0.17450980392156862, 1.0);
		rgb(18pt)=(0.0, 0.22156862745098038, 1.0);
		rgb(19pt)=(0.0, 0.25294117647058822, 1.0);
		rgb(20pt)=(0.0, 0.29999999999999999, 1.0);
		rgb(21pt)=(0.0, 0.33137254901960772, 1.0);
		rgb(22pt)=(0.0, 0.3784313725490196, 1.0);
		rgb(23pt)=(0.0, 0.40980392156862744, 1.0);
		rgb(24pt)=(0.0, 0.45686274509803909, 1.0);
		rgb(25pt)=(0.0, 0.50392156862745097, 1.0);
		rgb(26pt)=(0.0, 0.53529411764705859, 1.0);
		rgb(27pt)=(0.0, 0.58235294117647063, 1.0);
		rgb(28pt)=(0.0, 0.61372549019607847, 1.0);
		rgb(29pt)=(0.0, 0.66078431372548996, 1.0);
		rgb(30pt)=(0.0, 0.69215686274509802, 1.0);
		rgb(31pt)=(0.0, 0.73921568627450984, 1.0);
		rgb(32pt)=(0.0, 0.77058823529411768, 1.0);
		rgb(33pt)=(0.0, 0.81764705882352939, 1.0);
		rgb(34pt)=(0.0, 0.86470588235294121, 0.99620493358633777);
		rgb(35pt)=(0.0, 0.89607843137254906, 0.97090449082858954);
		rgb(36pt)=(0.034788108791903853, 0.94313725490196076, 0.93295382669196714);
		rgb(37pt)=(0.060088551549652112, 0.97450980392156861, 0.9076533839342189);
		rgb(38pt)=(0.098039215686274495, 1.0, 0.8697027197975965);
		rgb(39pt)=(0.12333965844402275, 1.0, 0.84440227703984827);
		rgb(40pt)=(0.16129032258064513, 1.0, 0.80645161290322587);
		rgb(41pt)=(0.18659076533839339, 1.0, 0.78115117014547764);
		rgb(42pt)=(0.22454142947501579, 1.0, 0.74320050600885512);
		rgb(43pt)=(0.26249209361163817, 1.0, 0.70524984187223283);
		rgb(44pt)=(0.2877925363693864, 1.0, 0.67994939911448449);
		rgb(45pt)=(0.3257432005060088, 1.0, 0.6419987349778622);
		rgb(46pt)=(0.35104364326375709, 1.0, 0.61669829222011385);
		rgb(47pt)=(0.38899430740037944, 1.0, 0.57874762808349156);
		rgb(48pt)=(0.4142947501581275, 1.0, 0.55344718532574344);
		rgb(49pt)=(0.45224541429475007, 1.0, 0.51549652118912082);
		rgb(50pt)=(0.49019607843137247, 1.0, 0.47754585705249841);
		rgb(51pt)=(0.5154965211891207, 1.0, 0.45224541429475018);
		rgb(52pt)=(0.55344718532574311, 1.0, 0.41429475015812778);
		rgb(53pt)=(0.57874762808349134, 1.0, 0.38899430740037955);
		rgb(54pt)=(0.61669829222011374, 1.0, 0.35104364326375714);
		rgb(55pt)=(0.64199873497786197, 1.0, 0.32574320050600891);
		rgb(56pt)=(0.67994939911448438, 1.0, 0.28779253636938651);
		rgb(57pt)=(0.70524984187223261, 1.0, 0.26249209361163817);
		rgb(58pt)=(0.74320050600885468, 1.0, 0.22454142947501621);
		rgb(59pt)=(0.78115117014547741, 1.0, 0.18659076533839347);
		rgb(60pt)=(0.80645161290322565, 1.0, 0.16129032258064513);
		rgb(61pt)=(0.84440227703984805, 1.0, 0.12333965844402273);
		rgb(62pt)=(0.86970271979759628, 1.0, 0.098039215686274495);
		rgb(63pt)=(0.90765338393421868, 1.0, 0.060088551549652092);
		rgb(64pt)=(0.93295382669196703, 1.0, 0.03478810879190386);
		rgb(65pt)=(0.97090449082858932, 0.95933188090050858, 0.0);
		rgb(66pt)=(0.99620493358633766, 0.93028322440087174, 0.0);
		rgb(67pt)=(1.0, 0.88671023965141638, 0.0);
		rgb(68pt)=(1.0, 0.84313725490196101, 0.0);
		rgb(69pt)=(1.0, 0.81408859840232406, 0.0);
		rgb(70pt)=(1.0, 0.7705156136528688, 0.0);
		rgb(71pt)=(1.0, 0.74146695715323196, 0.0);
		rgb(72pt)=(1.0, 0.69789397240377649, 0.0);
		rgb(73pt)=(1.0, 0.66884531590413965, 0.0);
		rgb(74pt)=(1.0, 0.62527233115468439, 0.0);
		rgb(75pt)=(1.0, 0.58169934640522891, 0.0);
		rgb(76pt)=(1.0, 0.55265068990559207, 0.0);
		rgb(77pt)=(1.0, 0.50907770515613682, 0.0);
		rgb(78pt)=(1.0, 0.48002904865649987, 0.0);
		rgb(79pt)=(1.0, 0.4364560639070445, 0.0);
		rgb(80pt)=(1.0, 0.40740740740740755, 0.0);
		rgb(81pt)=(1.0, 0.36383442265795229, 0.0);
		rgb(82pt)=(1.0, 0.33478576615831535, 0.0);
		rgb(83pt)=(1.0, 0.29121278140886042, 0.0);
		rgb(84pt)=(1.0, 0.24763979665940472, 0.0);
		rgb(85pt)=(1.0, 0.21859114015976777, 0.0);
		rgb(86pt)=(1.0, 0.1750181554103124, 0.0);
		rgb(87pt)=(1.0, 0.14596949891067557, 0.0);
		rgb(88pt)=(1.0, 0.1023965141612202, 0.0);
		rgb(89pt)=(0.99910873440285231, 0.07334785766158336, 0.0);
		rgb(90pt)=(0.94563279857397531, 0.029774872912127992, 0.0);
		rgb(91pt)=(0.90998217468805731, 0.00072621641249104307, 0.0);
		rgb(92pt)=(0.8565062388591802, 0.0, 0.0);
		rgb(93pt)=(0.80303030303030321, 0.0, 0.0);
		rgb(94pt)=(0.76737967914438521, 0.0, 0.0);
		rgb(95pt)=(0.71390374331550821, 0.0, 0.0);
		rgb(96pt)=(0.6782531194295901, 0.0, 0.0);
		rgb(97pt)=(0.6247771836007131, 0.0, 0.0);
		rgb(98pt)=(0.589126559714795, 0.0, 0.0);
		rgb(99pt)=(0.535650623885918, 0.0, 0.0);
		rgb(100pt)=(0.5, 0.0, 0.0);
}}

\DeclareSIUnit\dBm{dBm}
\DeclareSIUnit{\samplespersecond}{SPS}

\usepackage[bookmarks=true,breaklinks=true,colorlinks,linkcolor=black,citecolor=blue,urlcolor=black,pdfpagelabels=false]{hyperref} 
\usepackage[nameinlink,capitalise]{cleveref}

\paperheight=\pdfpageheight
\paperwidth=\pdfpagewidth

\fancypagestyle{firstpage}{
}

\title{Accelerating complex control schemes on a heterogeneous MPSoC platform for quantum computing}

\author{
\alignauthor
Richard Gebauer, Nick Karcher, Jonas Hurst, Marc Weber, and Oliver Sander\\
    \affaddr{Institute for Data Processing and Electronics, Karlsruhe Institute of Technology (KIT), Karlsruhe, Germany}\\
    \email{richard.gebauer@kit.edu}
}

\begin{document}
\maketitle
\thispagestyle{firstpage}
\pagestyle{plain}




\begin{abstract}
Control and readout of superconducting quantum bits (qubits) require microwave pulses with gigahertz frequencies and nanosecond precision.
To generate and analyze these microwave pulses, we developed a versatile FPGA-based electronics platform.
While basic functionality is directly handled within the FPGA, guaranteeing highest accuracy on the nanosecond timescale, more complex control schemes render impractical to implement in hardware.

To provide deterministic timing and low latency with high flexibility, we developed the Taskrunner framework.
It enables the execution of complex control schemes, so-called user tasks, on the real-time processing unit (RPU) of a heterogeneous Multiprocessor System-on-Chip (MPSoC).
These user tasks are specified conveniently using standard C language and are compiled automatically by the MPSoC platform when loaded onto the RPU.
We present the architecture of the Taskrunner framework as well as timing benchmarks and discuss applications in the field of quantum computing.
\end{abstract}

\section{Introduction}
Quantum bits (qubits) are the elementary building blocks of a quantum processor. While a variety of different qubit implementations exist \cite{Monroe1164IonQC,PhotonQC,SemiconductorQC}, superconducting circuits represent a promising candidate that is actively used and investigated by many groups in the field \cite{TransmonKoch,Devoret1169,QuEngGuideScQubits}.
In the last couple of years, the focus shifted from mostly physics-driven research focusing on qubit types and properties towards a more comprehensive, interdisciplinary system perspective \cite{Devoret1169OutlookScQub,ScalableScQubSurfaceCode,GoogleQuantumSupremacy}. In order to build a quantum computer, the full software and hardware stack \cite{QCStack, QCStack2, QMicroArchitecture} must be considered. One of the key requirements is a quantum-classical interface to control the quantum processor.

Systems based on superconducting qubits \cite{TransmonKoch,Devoret1169,QuEngGuideScQubits} require microwave pulses with gigahertz frequencies and nanosecond precision for readout and control \cite{DispReadout,QNDSuperconducting,PRL2005ScQubControl}.
A quantum-classical interface thus needs to be able to transform classical instructions into such microwave pulses that then interact with the qubits.
Furthermore, the response of the setup needs to be monitored and analyzed in order to determine the qubit states.
We developed a versatile electronics platform based on a field-programmable gate array (FPGA) that provides such an interface for superconducting qubits.
The most important functionalities, like scheduling microwave pulses with nanosecond accuracy and evaluating readout responses, is directly implemented on the FPGA.
Control schemes that require challenging parameter variations and advanced data evaluation are implemented in software.
This combines the advantages of a versatile and fast software development process with the benefits of highly parallel data processing and nanosecond precision on the FPGA.

We utilize a heterogeneous system architecture based on the Xilinx Zynq UltraScale+ series which combines the FPGA with an ARM Cortex-A53 application processor (APU) and an ARM Cortex-R5 real-time co-processor (RPU). By providing a software running on the RPU called Taskrunner, the user can easily implement and load complex experiment schemes during run-time. This is complemented by on-the-fly compilation of the user task on the APU. The RPU is closely connected with the FPGA domain which enables it to quickly and deterministically access hardware registers. It also provides the ability to perform further data reductions in software.

In the following, we will give a brief introduction into the fundamentals of interfacing with superconducting qubits and related work in the field. This is complemented by an introduction into the used hardware, our platform architecture and the interaction between the components.
In the main chapter of this work, we present the Taskrunner framework.
After elaborating on the architecture, we give insights into different aspects of the Taskrunner.
This ranges from the initialization process over loading and executing a user task up to the interface used to exchange information and data with the external client.
Finally, we provide timing benchmarks for the most common use-cases and present selected applications in experiments with superconducting qubits.

\section{Fundamentals}

\subsection{Interfacing superconducting qubits}\label{sec:qubit-control}
Qubits are artificial atoms that can be interacted with by employing light-matter interaction \cite{SchoenShnirmanQubitStateEngineering, TransmonKoch}.
Therefore, microwave photons with frequencies of a few gigahertz are regularly used to manipulate and readout the state of superconducting qubits \cite{QuEngGuideScQubits, DispReadout, PRL2005ScQubControl}.

Qubits are operated fundamentally different than classical bits. 
While information of classical bits is represented as electrical voltages, quantum bits have a quantum mechanical state comprising two fundamental states called $\ket{0}$ and $\ket{1}$. 
In contrast to classical bits, arbitrary superpositions of these two states are possible and can be exploited for computations \cite{QITBennettShort98}.

\subsubsection{Qubit gate operations}
For most qubit implementations, gates are also realized differently than on a classical computer.
There, gates are stationary circuits built out of transistors through which the bits are passed through in form of electrical signals.
In quantum computing, especially for superconducting qubits, the qubits themselves are stationary and gate operations are performed by applying microwave and/or current pulses to the qubit circuitry. 
With typical frequencies of multiple gigahertz and pulse durations of a few tens of nanosecond, high sampling quality and speed is necessary in order to achieve accurate qubit gates \cite{OptDrivingScAtomChow2010}.

Quantum computing algorithms comprise a sequence of gate operations and thus microwave pulses. Timing of these pulses is critical and the phase jitter can easily compromise the calculations \cite{DephasingInducedPhotonNoiseBertet05, RandBenchProcTomGateErrorsChow09}.

\subsubsection{Qubit state measurements}
Qubit state measurements play a central role in every quantum algorithm, be it to obtain the result at the end of the calculation or to influence the algorithm through measurement back action \cite{QuantumBackactionHatridge13}.
In quantum computing, measuring qubits will influence their state after the measurement.
While, during calculation, the qubit can be in any arbitrary superposition, the readout will result in either the $\ket{0}$ or $\ket{1}$ state.
The probabilities of the two possible measurement outcomes are given by the superposition.
After the measurement, the superposition is destroyed and the qubit is in the detected state.
This is called a projective measurement and characteristic for experiments in quantum mechanics \cite{NoCloningQMPark1970}.

This also implies that, in order to obtain the probabilities, one cannot simply repeat the measurement operation numerous times.
Instead, one needs to repeat the whole pulse sequence to end up with the same superposition prior to the projective measurement.
For quantum computing, one therefore tries to formulate the algorithm in a way that, prior to the measurement of the final result, the qubit is in one of the two states $\ket{0}$ or $\ket{1}$.

\subsubsection{Parameter variations and averaging}
In research on superconducting qubits, on the other hand, the probabilities are of great interest.
Therefore, many repetitions of the pulse sequences will be performed in order to obtain suitable statistics.
This is often combined with parameter variations to investigate the qubit state dependency on these parameters.
The order of averaging and parameter variations can have a large impact on the measurement result if low frequency noise is present in the system.

When first averaging for one parameter set, a currently present noise bias will be incorporated into the signal.
For the next parameter configuration, the bias might have changed if the noise frequency is on the order of the parameter change rate.
To give an example:
A single pulse sequence takes \SI{10}{\micro\second} and we perform \num{1000} averages per parameter configuration. 
That implies a parameter change every \SI{10}{\milli\second} or a rate of \SI{100}{\hertz}.
Thus, it makes the experiment outcome highly sensitive to noise on this time scale, like \SI{50}{\hertz} noise.
It is therefore generally advisable to first change the parameters and average afterwards.
While this is conceptually easy, depending on the experimental setup, it might be difficult to implement efficiently.

\subsubsection{Qubit relaxation delay}
In order to obtain reproducible results, it is furthermore important that the qubits are all in the same state before each pulse sequence starts.
In the most simplest case, this is achieved by waiting a sufficiently large time interval after the last sequence.
The qubit is very fragile and its state decays exponentially into the $\ket{0}$ state on a time scale $T_1$, called energy relaxation time.
Typical $T_1$ times for superconducting qubits range from $1 - \SI{100}{\micro\second}$ \cite{TransmonKoch, ScQubit100musRigetti2012}.
Therefore, one typically waits on the order of five to ten times the $T_1$ time to ensure that the qubit is relaxed into $\ket{0}$ and one has comparable starting conditions.

\subsubsection{Dispersive qubit readout}
Returning to the measurement itself, this is normally performed as so-called dispersive readout \cite{DispReadout}.
In that case, the qubit is coupled to a microwave resonator leading to a slight shift of the resonator frequency depending on the qubit state.
By probing the microwave resonator with a readout pulse, typically a few hundred nanoseconds long, one can extract the qubit state.
It is encoded in the phase information of the reflected microwave pulse.
By performing heterodyne IQ mixing and a digital down-conversion, the in-phase and quadrature components of the signal can be obtained, and thereby the phase information.
In \cref{fig:iqclouds-transmon}, the measurement result of an actual qubit experiment is shown.

\subsection{Related work}
Research institutions working with superconducting qubits widely use generic laboratory equipment to generate and analyze the necessary microwave pulses.
This includes arbitrary waveform generators (AWGs), vector network analyzers (VNAs) and signal analyzers (SAs).
They are suited for fast testing of circuits and performing experiments with simple pulse sequences.
With increasing system complexity, utilizing these devices becomes unfeasible due to significant communication delays, bad scaling properties and high relative cost.
Complex data processing like calculating correlations is also inefficient or even impossible to realize with such setups.
Therefore, FPGA-based systems emerged in different research groups specifically designed for certain experiments to meet the high data processing and latency demands \cite{RisteFeedbackFPGA,OfekLifetimeExtensioFPGA,EntanglementStabilizationFPGA}.

Recently, first commercial products appeared on the market specifically targeting the quantum computing sector for superconducting qubits. Noteworthy products are the OPX of Quantum Machines \cite{QuantumMachines}, the Quantum Computing Control System (QCCS) of Zurich Instruments \cite{ZurichInstrumentsQCCS}, and the Quantum Engineering Toolkit (QET) of Keysight \cite{KeysightQET}. All offer the sequencing, generation and detection of base-band microwave pulses. The technical realization, however, differs among the different products.

Keysight's QET comprises a modular system in a PXIe chassis with components to generate microwave signals, components to digitize and process microwave pulses, and components to realize the control of the other components. The AWG components each offer four \SI{14}{\bit} channel outputs with \SI{1}{\giga\samplespersecond} or \SI{16}{\bit} outputs with \SI{500}{\mega\samplespersecond}. The digitizer components offer up to eight \SI{14}{\bit} input channels with \SI{100}{\mega\samplespersecond} or four \SI{14}{\bit} channel with \SI{500}{\mega\samplespersecond}. 
The modules are based on FPGAs of the Xilinx Kintex-7 series.

Similarly, Zurich Instruments' QCCS builds on a combination of their HDAWG devices for manipulation pulse generation and a separate device to generate and process readout pulses, the UHFQA Quantum Analyzer.
The HDAWG offers up to eight \SI{16}{\bit} channels with \SI{2.4}{\giga\samplespersecond}. The UHFQA has a \SI{14}{\bit} dual-channel AWG and a \SI{12}{\bit} dual-channel input, both operating at \SI{1.8}{\giga\samplespersecond}. They offer frequency-division multiplexing for up to ten qubit readout frequencies.
In contrast to Keysight's product, these are fully operational stand-alone devices.
Multi-device synchronization is achieved by an additional device, the Programmable Quantum System Controller (PQSC). There, they use a Xilinx UltraScale+ XCZU15EG-2I FPGA \cite{ZurichInstrumentsQCCS} with heterogeneous MPSoC that can be customized by the user. It is not publicly known if the real-time processor is used and if a similar FPGA is present within UHFQA and HDAWG.

Quantum Machine's OPX integrates pulse generation and qubit readout in a single device. It offers ten analog output channels and two analog input channels.
To the best of our knowledge, it does not utilize a heterogeneous MPSoC processing system.

Besides the different quantum readout systems, there are theoretical concepts to describe future quantum computers with a higher level of abstraction.
A quantum computing stack \cite{QCStack, QCStack2, QMicroArchitecture} describes the necessary layers to translate a high-level description of a quantum algorithm into physical operations on a quantum processor.
The aforementioned hardware platforms provide the lowest layer above the quantum chip itself and act as quantum-classical interface.
If their classical computation capabilities permit, they can also be utilized to implement a simple quantum computer micro-architecture.

\section{Platform architecture}
\begin{figure}
    \centering
    \includegraphics[width=0.4\textwidth]{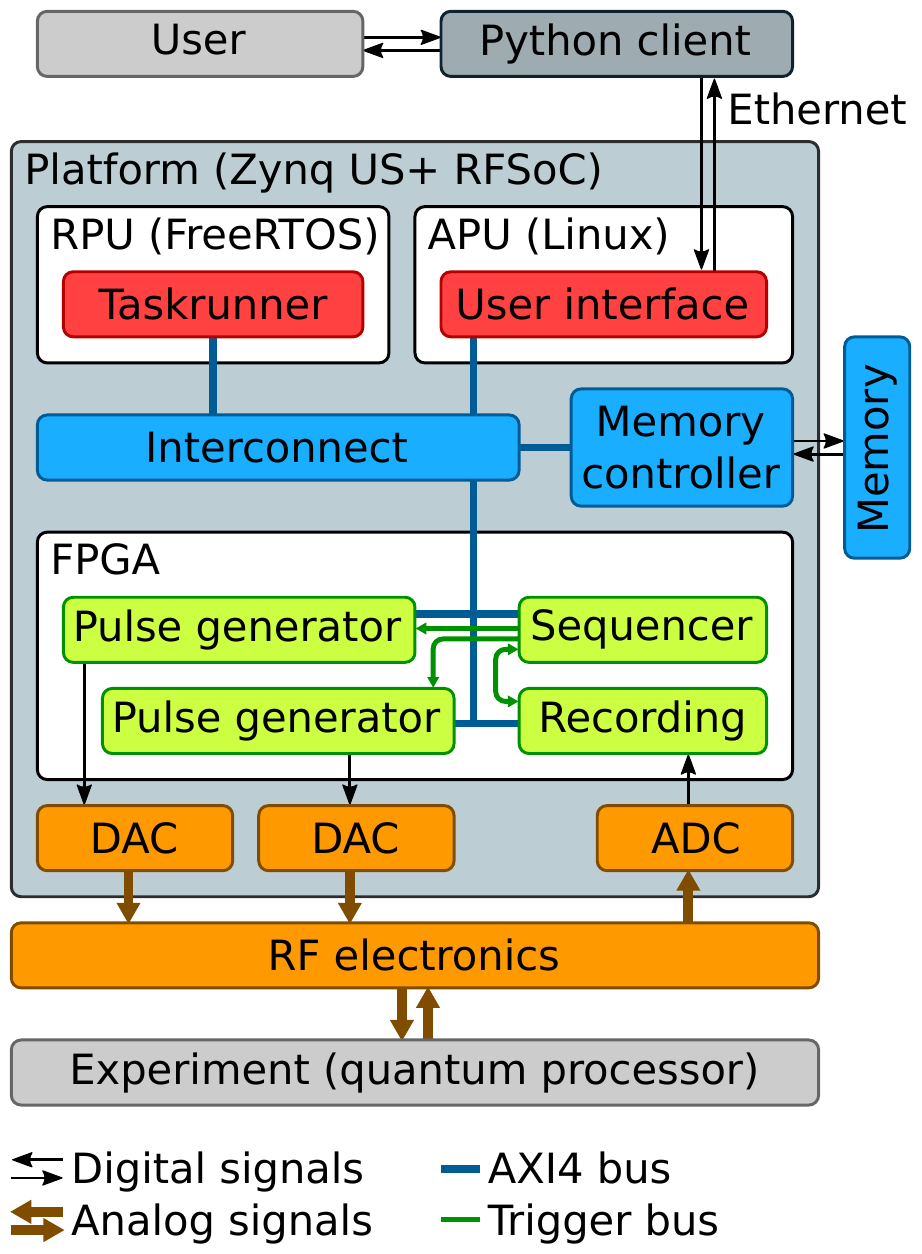}
    \caption{Architecture of our heterogeneous electronics platform.}
    \label{fig:platform-overview}
\end{figure}
Our electronics platform is based on the heterogeneous architecture provided by Xilinx Zynq UltraScale+ MPSoCs.
After first using the ZCU102 evaluation board with externally attached FMC converter cards, we changed our focus to the ZCU111 board.
It features the more integrated RFSoC comprising gigasample A/D and D/A converters directly integrated within the system-on-chip (SoC).
An overview of our design architecture is visible in \cref{fig:platform-overview}.
To generate and digitize microwave pulses, we employ the integrated A/D and D/A converters operating at \SI{4}{\giga\samplespersecond} and utilize decimation and interpolation filters to obtain a per-channel data rate of \SI{1}{\giga\samplespersecond} within the FPGA.
This allows the platform to handle base-band signals up to the Nyquist frequency of $f_s /2 = \SI{500}{\mega\hertz}$.
A separate RF electronics containing mixers and microwave sources is necessary to translate these base-band signals to and from the band-pass signal in the desired gigahertz frequency range of the superconducting circuits.

On the FPGA, different modules are implemented.
\cref{fig:platform-overview} shows the modules needed to interact with a single qubit. 
Two pulse generators supply the DACs with samples containing digital pulses.
Frequency, phase, and shape of these pulses can be defined during the experiment.
One generator is used for manipulating the qubit state, the other to generate readout pulses.
A recording module processes the experiment response from these pulses by performing a digital down-conversion of the signal obtained from the ADC.
It also applies further filtering and processing to average over measurements and extract the state of the qubit.
Both pulse generators and the recording module are triggered by a sequencer that enables the user to schedule pulses in \SI{4}{\nano\second} steps, as well as to start processing of incoming pulses.
The recording module also reports the determined qubit states back to the sequencer which can then perform sequences conditioned on the outcome of a previous measurement.
All modules are supplied by a single clock domain rigidly coupled to the converter clock.
This guarantees highest reproducibility between multiple experiment repetitions.

All modules are addressable by a register-based AXI4Lite bus and mapped into the physical memory address range of the processing system.
Therefore, RPU and APU can read out status and data from these modules, as well as configure them, by simple memory read and write operations.
They also share a common external DDR4 memory which can be used to exchange data between both processors.

The APU hosts a Linux operating system on which programs are implemented to initialize the platform at boot time, provide means to communicate with external clients via Ethernet, as well as to perform further data post-processing online.
We wrote a modular communication server running on the APU with services for all of our components.
This includes FPGA components like the pulse generators, the sequencer and the recording module, but also more abstract components like the Taskrunner.
The communication protocol is based on remote procedure calls (RPC) and employs the open source framework gRPC \cite{gRPC}.
Thereby, it is possible to connect to the platform with any client written in a language that is supported by gRPC.
As Python is widely used within the experimental physics community, this was our choice for the primary client.

\section{Taskrunner framework}
As detailed in \cref{sec:qubit-control}, performing computational tasks and experiments with superconducting qubits requires the execution of a well-defined sequence of pulses and measurements. Typically, delays and pulse properties need to be adapted between individual executions to sweep a specified parameter range.
Executions with fixed properties are directly handled by the sequencer within the FPGA for highest reproducibility with nanosecond precision.
However, performing complex sequences and varying parameters between individual ones is impractical to implement in hardware.
Implementing these variations in the Python client is easily possible but yields a high latency due to the communication overhead.

The main goal of our Taskrunner framework is to address this issue by utilizing the real-time co-processor (RPU) on the heterogeneous Zynq UltraScale+ architecture.
This renders complex control schemes with deterministic timing and low latency possible without having to adapt the FPGA.
The Taskrunner framework enables the user to execute arbitrary code, so-called user tasks, at run-time.
Tasks can be transmitted in binary format or as source code.
In the latter case, the source code will be compiled on the fly by the APU before being loaded onto the Taskrunner.
No external compiler is needed on the user computer.
This further reduces external dependencies and makes it easy for users to facilitate this subsystem.
The Taskrunner provides a standardized interface to interact with running tasks and exchange information.
It also reduces technical complexity and accelerates the development and experimental testing process.

\subsection{Architecture}
\begin{figure}
    \centering
    \includegraphics[width=0.42\textwidth]{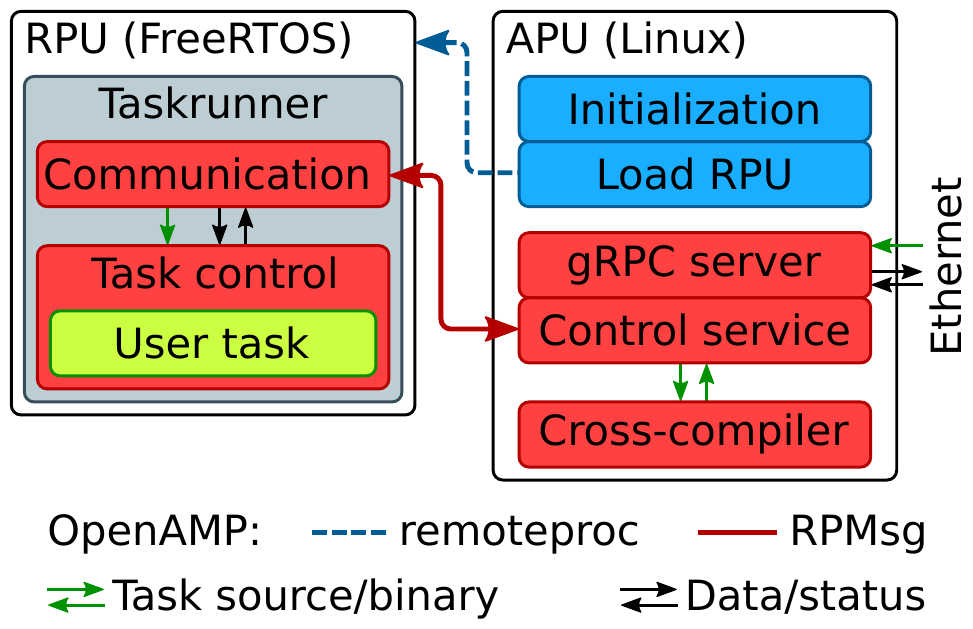}
    \caption{Structure of the Taskrunner framework on RPU and APU.}
    \label{fig:taskrunner-overview}
\end{figure}
The structure of the Taskrunner framework is depicted in \cref{fig:taskrunner-overview}.
It comprises the APU and RPU, as well as the communication between both processors and the communication to an external client via Ethernet.

\subsubsection{Real-time co-processor}
On the RPU, our Taskrunner software is hosted by the real-time operating system FreeRTOS.
It handles communication with the APU and controls the task execution and is loaded together with FreeRTOS upon initialization from the APU.

The Taskrunner consists of two threads that are scheduled by the FreeRTOS scheduler. One handles the communication with the control service on the APU. It is only scheduled when the control service is requesting information from or sending data to the Taskrunner. The other thread implements the execution of the user task, in the following called application thread. Unwanted context switches between both threads that would break deterministic execution in relevant sections of the user task can be inhibited by defining a critical section using the Taskrunner interface (see \cref{sec:TaskrunnerInterface}).

\subsubsection{Application processor}
Most of the functionality provided by the Taskrunner is directly wrapped by the control service on the APU and exposed to the user.
The functionality also includes loading of new tasks during run-time which, if transmitted as source code, will first be compiled on the fly by a cross-compiler on the APU and afterwards sent to the Taskrunner. There, the new task replaces a possibly pre-existing user task and can be started and stopped by the gRPC control service.

\subsubsection{Inter-processor communication}
The connection between RPU and APU employs the shared DDR4 memory, as well as the Linux remoteproc framework and the RPMsg messaging protocol \cite{OpenAMP_RPMsg_Github}.
On the RPU, the implementation is provided by the Open Asymmetric Multi Processing (OpenAMP) \cite{OpenAMP_Github} framework.
In our configuration, the APU acts as host, controlling the RPU slave via remoteproc. 
The communication between the Taskrunner on the RPU and the control service on the APU is handled by the RPMsg protocol.
RPMsg is based on inter-processor interrupts and a shared memory (compare \cref{fig:platform-overview}).
The communication thread on the RPU is notified by an interrupt whenever a message is available.
In the other direction, the control server on the APU is currently only polling status information on request.

Building on the RPMsg protocol, the user payload is further structured.
After a header to indicate message type and packet number with \num{1}\,byte each, the type-specific packet data follows.
We implemented the following packet types:
acknowledged, not acknowledged, status request, control operation like starting and stopping a user task, update parameter list size, obtain error messages in queue, get finished data boxes, mark data boxes as processed, set the Taskrunner firmware hash, as well as packets to transfer the user task and to init and close the connection.

\begin{listing}[t]
	\begin{minted}[
	fontsize=\footnotesize,
	breaklines,
	obeytabs=true,tabsize=2,
	linenos
	]{C}
#include "task.h"
#include "recmodule.h"
#include "sequencer.h"

int task_entry()
{
	uint32_t *param_list = rtos_GetParameters();
	uint32_t param_count = rtos_GetParametersSize() / sizeof(uint32_t);
	if (param_count != 2)
	{
		rtos_PrintfError("Please provide exactly 2 parameters (%d given).", param_count);
		return -1;
	}
	uint32_t repetitions = param_list[0];
	uint32_t start_pc = param_list[1];
	
	iq_pair *data_iq = rtos_GetDataBox(repetitions * sizeof(iq_pair));
	
	// Ensure sequencer is ready for current task
	sequencer_wait_while_busy();
	
	for (uint32_t i = 0; i < repetitions; i++)
	{
		sequencer_wait_until_qubit_relaxed();
		
		sequencer_start_at(start_pc);
		
		// Wait until result available
		sequencer_wait_while_busy();
		recmodule_wait_while_busy(0);
		
		recmodule_get_iq_pair(0, data_iq + i);
		
		rtos_SetProgress(i + 1);
	}
	
	rtos_FinishDataBox(data_iq);
	return 0;
}
	\end{minted}
	\caption{Basic user task showcasing how to implement a control flow with the Taskrunner framework. This task will perform a certain number of repetitions as defined in the parameter list and store every single in-phase and quadrature result in a data box. The obtained data can be used to generate a histogram as shown in \cref{fig:iqclouds-transmon}.}
	\label{listing:basic-iqclouds-task}
\end{listing}

\subsection{Initialization}
During boot of the Linux on the APU host, an init script initializes the RPU.
It first ensures that the RPU is stopped.
Then it copies the Taskrunner binary including FreeRTOS as firmware onto the RPU.
Both Taskrunner firmware and user task are located inside the tightly-coupled memory (TCM) banks of the Zynq UltraScale+ series \cite[p.~80~f.]{ZynqUSPlusRTU}.
This ensures a low-latent, deterministic access and thus execution of operations.
The TCM banks are operated in lock-step mode, giving access to the full \SI{256}{\kilo\byte} memory capacity.
To maximize performance, text sections are located inside the first TCM (ATCM) and data sections inside the second one (BTCM)\cite[p.~89]{ZynqUSPlusRTU}.
For the user task, a \SI{50}{\kilo\byte} region is reserved in both ATCM and BTCM.
The remaining space is utilized by the firmware itself, which is occupying \SI{123}{\kilo\byte} of memory.

At the end, the RPU is started and it is checked that everything is running.
If so, a Linux kernel module handling the communication with the RPU is loaded.
Finally, a MD5 hash of the Taskrunner firmware is sent to the Taskrunner which can later be queried in order to verify that pre-compiled user tasks are linked against the correct Taskrunner version.
Afterwards, the gRPC server is started and the control service initialized.
The service establishes the RPMsg connection to the Taskrunner and queries the memory regions for data and parameter transfer.

\subsection{User Task Programming}
The user task is conveniently written in the C language. Thereby, a wide feature set of the C language and existing libraries can be utilized.
A basic task triggering pulse sequences and collecting measurement results is shown in \cref{listing:basic-iqclouds-task}.
The entry function of the user task is called \mintinline{C}{task_entry()} and has no parameters. All interaction with the Taskrunner is handled by separate interface functions (see \cref{sec:TaskrunnerInterface}). 

The source code is transmitted to the control service on the APU via the gRPC interface. There, a cross-compiler for the RPU is invoked.
In our case, this is the bare-metal compiler of the GCC (\texttt{arm-none-eabi}).
We also provide multiple C libraries to abstract from a pure register-based access when interacting with modules on the FPGA. This e.g. includes functions like \mintinline{C}{sequencer_wait_while_busy()}. They are already present on the platform and will be taken into account when the source file is compiled.
Furthermore, in order to interact with the Taskrunner and provide standard libraries, the user task is also linked against the RPU firmware binary.
It is noteworthy that the compilation is performed with the optimization flags \texttt{-O2 -mcpu=cortex-r5 -funroll-loops} which substantially speeds up the task execution.
After compilation, the binary is stripped using the \texttt{objcopy} method of the compiler removing unused sections in the binary.
A custom linker script, ensures that the entry function is at the beginning of the resulting binary file and the function symbol addresses are matched to the running FreeRTOS.

Finally, meta information is added to the end of the binary file, in particular a configurable name for the task as well as the MD5 hash of the RPU firmware binary. The latter is to guarantee compatibility in case the binary is kept and reused later. It will be checked by the logic loading the user task and results in a rejection if it does not match the running version.
If an error occurs during the compilation process, the output streams of the compiler will be returned to the user.
The user can also load a pre-compiled task via the gRPC interface which will then skip the compilation step. In both cases, the MD5 hash of the task and the running RPU firmware is compared to ensure compatibility before the binary is loaded onto the RPU using the RPMsg interface.

On the RPU, the communication thread will store the task in the TCM. 
Both the communication and the application thread of the Taskrunner use a mutex to guarantee that there is no conflicting access to the user task memory region.
Direct access from the APU to the TCM is not intended.
After loading the user task, the status flags are updated accordingly.

\subsection{User Task Execution}
Prior to starting the task execution, parameter values can be passed to the user task.
Examples are a list of delays, the number of average repetitions to perform, or boolean information like if a separate calibration step should be included in the execution.
The parameter list will be transferred via gRPC to the APU and copied into the respective region in the DRAM by the control service.
In our implementation, the region currently offers \SI{15}{\mega\byte} of space for parameter values.
The control service also notifies the Taskrunner that the parameter list has been updated and reports the new size of the list.
This information can later be queried by the user task to know the valid parameter range inside the DRAM memory region.
As the DRAM access is subject to fluctuations in execution time, the on-chip memory (OCM) \cite[p.~35]{ZynqUSPlusRTU} could potentially also be used.
With its size of \SI{256}{\kilo\byte} it should be still sufficient for most applications. 

When the user starts the task execution, the command will be relayed via gRPC and RPMsg to the Taskrunner.
The communication thread of the Taskrunner will ensure that no task is currently running and raise an error if it is the case.
Then it will notify the application thread about the received start command.
The application thread will try to lock the mutex of the user task memory region.
It then verifies that a user task was loaded and a start command issued.
Before starting the task, it invalidates the DRAM cache of the region where the parameter values are located.
This is especially important if the values changed since the last execution to avoid obtaining outdated parameter values.

It then casts the address of the task memory region into a function pointer and calls the entry function of the user task in the context of the application thread.
After the execution finished, the mutex is released again.
Before and after the function call, also the status flags are updated accordingly so the communication thread reports the correct status of the Taskrunner.

\subsection{Taskrunner Interface}\label{sec:TaskrunnerInterface}
The user task can utilize a provided function library to interact with the Taskrunner. The functions are part of the RPU firmware binary and will be linked against during compilation. A list of all available functions is given in \cref{listing:taskrunner-interface}. This includes a print function to output text to the UART interface for debugging purposes and functions wrapping FreeRTOS functionality like defining critical sections and resetting and reading out the PMCCNTR counter for timing benchmarks.
Error messages can be specified and will be appended to a queue that can be fetched by the control service on the APU.
The Taskrunner also provides a function to query the parameter list that can be used to customize the task's behaviour.
One function allows the user task to specify a \mintinline{c}{uint32_t} progress value that can be polled by the user during execution to monitor the task progress.
Finally, functions to operate with data boxes (see next section) are provided to exchange data with the user.
\begin{listing}[b]
	\begin{minted}[
	fontsize=\footnotesize,
	breaklines,
	obeytabs=true,tabsize=2
	]{C}
void rtos_printf(const char *format, ...);
void rtos_EnterCriticalSection(void);
void rtos_ExitCriticalSection(void);
void rtos_RestartTimer(void);
uint32_t rtos_GetCycleCountTimer(void);
uint32_t rtos_GetNsTimer(void);
void rtos_ReportError(const char *error_msg);
void rtos_PrintfError(const char *format, ...);
void *rtos_GetParameters(void);
uint32_t rtos_GetParametersSize(void);
void rtos_SetProgress(uint32_t progress);
void *rtos_GetDataBox(uint32_t size);
void rtos_FinishDataBox(void* databox);
void rtos_DiscardDataBox(void* databox);
	\end{minted}
	\caption{A list of all functions available for the Taskrunner interface.}
	\label{listing:taskrunner-interface}
\end{listing}

\subsection{Data Transfer}
Typical user tasks in experiments with superconducting qubits perform data collection, aggregation or post-processing.
Therefore, the Taskrunner provides an interface to exchange collected data with the APU and the user.
In our implementation, we employ so-called data boxes to store obtained or processed values from the user task.
We define a \SI{480}{\mega\byte} heap located in the DDR4 memory with custom functions to request, finish and discard data boxes of an individually settable size (see last section).

Finished data boxes of a running or completed user task can be requested by the user.
The control service then fetches a list of the corresponding DDR4 memory addresses and sizes from the Taskrunner.
After sending the data boxes to the user, the corresponding memory regions are freed and can be reused again.
When discarding an acquired data box the allocated memory region is also freed.
In this case no data is sent to the APU and user.
Data boxes from a previous task execution which have not been discarded or fetched will be freed when a new task starts.

When transferring data between Taskrunner and user, two different operation schemes can be distinguished.
For shorter experiments it is sufficient to collect the data when the user task completed.
The data boxes are finished at the end of the user task and the Python driver checks if the task is done prior to fetching the data boxes.
During task execution, only the progress value is exchanged between Taskrunner and client.

Longer experiments, on the contrary, might exceed data sizes that can be handled by the heap on the RPU.
Furthermore, fast feedback by delivering first data to the user is beneficial as they might already judge from this if the task should continue or if some experimental parameters need to be adapted.
Therefore, data boxes can also be finished during the user task execution and collected in parallel by the user.
Afterwards, the memory is freed again on the RPU which solves the first issue.
It also provides the means to directly visualize the evolving measurement results while the experiment progresses.

\section{Performance validation}
In order to verify operation of the system, the following section presents characteristic timing benchmarks and elaborates on means for improvement.
It also shows selected experimental applications and results to showcase operation scenarios in superconducting qubit experiments.

\subsection{Timing benchmarks}
During execution of the user task in the Taskrunner, each communication request causes a context switch between the two FreeRTOS threads and thus an interruption of the user task.
Depending on the complexity of the request, the interruption can be shorter or longer.
Requesting the Taskrunner status and obtaining the newest progress value, for example, causes an interruption of \SI{16,2}{\micro\second}.
Querying error messages when none are present leads to a \SI{14,3}{\micro\second} interruption.
Asking for available finished data boxes takes longer and causes an interruption of \SI{42,7}{\micro\second}.
Therefore, it is advisable to query the task status not too often in order to not significantly slow down the user task execution.
In our Python client, we typically wait \SI{200}{\milli\second} before checking again if the user task completed.
This is to minimize the impact of the interruptions due to status requests and still get a reasonably fast reaction.
Also, critical sections should be defined where necessary.

\begin{table}
    \centering
    {\fontsize{9pt}{10.8pt}\selectfont
    \begin{tabular}{r | r r}
        Operation & Taskrunner & Python client \\
        \hline
        AXI register read & \SI{306 \pm 2}{\nano\second} & \SI{590\pm40}{\micro\second} \\ 
        AXI register write & \SI{323 \pm 2}{\nano\second} & \SI{620\pm40}{\micro\second} \\ 
        Sequencer status poll & \SI{324 \pm 2}{\nano\second} & \SI{580\pm60}{\micro\second} \\ 
        AXI reg. memcpy & \SI{312401\pm4}{\nano\second} & \SI{1500\pm130}{\micro\second} \\ 
        Array multiplication & \SI{10270}{\nano\second} & \SI{1,5\pm0,8}{\micro\second} \\ 
    \end{tabular}
    }
    \caption{Comparison of timing performance for typical operations running within the Taskrunner and within the Python client. The AXI register memcpy was benchmarked with 1024 register values copied into the TCM. The array multiplication was performed with two 32\,bit arrays, each with 1024 elements, also located in the TCM.}
    \label{tab:operation-timing}
\end{table}
Apart from interruptions caused by the communication between the processors, execution times of typical operations on the Taskrunner are of interest.
These are shown in \cref{tab:operation-timing}.
To illustrate the numbers, the operations running on the Taskrunner are compared to a version where they are executed by the Python client and handled by the gRPC server.
For most operations, the Taskrunner is significantly faster due to the more direct connection to the FPGA modules.
Only for the array multiplication (\num{1024} elements), the Python client is faster as it can leverage on a stronger desktop processor and the Python numpy library.
With except to the first two executions, which took \SI{14}{\nano\second} longer, we observed constant execution time of the array multiplication on the Taskrunner.
This indicates deterministic behavior as expected when only RPU and TCM memory are involved.
We verified that the increased duration of the first two executions is caused by the branch prediction feature of the ARM Cortex-R5 processing unit in the loop used for the calculation.
\num{100000} repetitive measurements of each operation have been performed to obtain the presented average durations and standard deviations.

\begin{table}
    \centering
    {\fontsize{9pt}{10.8pt}\selectfont
    \begin{tabular}{r r | r r}
        Task & Lines & Source code & Binary file \\
        \hline
        Empty & \num{6} & \SI{184,8 \pm 0,5}{\milli\second} & \SI{1,25 \pm 0,09}{\milli\second} \\ 
        Basic & \num{39} & \SI{260,3 \pm 0,6}{\milli\second} & \SI{1,55 \pm 0,10}{\milli\second} \\ 
        Complex & \num{386} & \SI{1107,2 \pm 1,8}{\milli\second} & \SI{2,99 \pm 0,10}{\milli\second} \\ 
    \end{tabular}
    }
    \caption{Durations to load different tasks onto the Taskrunner from the Python client. Three tasks are distinguished based on their complexity which is also reflected in the number of source code lines.}
    \label{tab:taskloading-timing}
\end{table}
Another figure of merit is the duration to load a task onto the Taskrunner from the Python client.
Timing results for different tasks are presented in \cref{tab:taskloading-timing}.
The durations to load the source code are each averaged over \num{100} iterations while the durations to transfer the binary file, due to the lower values and thus higher relative fluctuations, are averaged over \num{10000} repetitions.
In all cases, the duration to load the source file is dominated by the compile time as loading the precompiled binary is significantly faster.
The increased compile time is due to the use of optimization flags.
It is therefore justified as it will speed up the user task execution.
The duration to load a source task also shows a clear dependency on the complexity of the task.
While the empty and basic task both only consist of the entry function, the complex task also implements multiple separate functions to perform calculations like an FFT.

Finally, the speed to transfer data boxes to the Python client is an important benchmark.
Requesting and transferring \SI{10}{\kilo\byte} data as \mintinline{C}{int32_t} values takes \SI{4,3 \pm 1,6}{\milli\second}.
With increasing data box size, the overhead decreases and thus the transfer speed increases.
Transferring a \SI{100}{\mega\byte} data box requires \SI{2,32 \pm 0,16}{\second} corresponding to \SI{43}{\mega\byte\per\second}.

\subsection{Experimental applications}
In the field of quantum computing, many applications exist where a low response time is critical but prototyping an FPGA design is unfeasible.
While not being discussed in this paper, this can also comprise tasks like prototyping complex quantum error correction schemes or realizing part of a quantum processor micro-architecture.
As these concepts are still at an early stage, in the following, we will instead focus on three fundamental use-cases in research with superconducting qubits.

\subsubsection{Fast parameter changes}
As outlined in \cref{sec:qubit-control}, it is important to first change parameter values before averaging in order to avoid low frequency noise.
For operation without the Taskrunner, this requires frequent communication between the FPGA and our external Python client.
Let us assume a standard experiment with \num{42} parameter changes, a qubit relaxation delay of \SI{100}{\micro\second} and \num{10000} repetitions.
This would roughly take \SI{24}{\hour} due to the significant communication overhead after each pulse sequence.
In contrast, when first averaging on the sequencer and then changing the parameters using Python, this example experiment takes \SI{51}{\second} but becomes susceptible to low frequency noise. 

At the same time, an identical experiment can be easily implemented using the Taskrunner framework and executed in only \SI{43}{\second}.
In this case, a user task performs the necessary parameter changes, collects single measurement results and only afterwards averages them separately for each parameter configuration.
This solves both the overhead issue and the susceptibility to slow fluctuating noise.
The major contribution in both cases is the qubit relaxation delay corresponding to $\num{10000} \times \num{42} \times \SI{100}{\micro\second} = \SI{42}{\second}$.
The difference is that, when executed by the Taskrunner, there is nearly no overhead left due to the low-latent interface towards the FPGA.

\subsubsection{Single measurement statistics}
\begin{figure}
    \centering
    \begin{tikzpicture}
    \begin{axis}[
    width=0.36\textwidth, height=0.33\textwidth,
    colorbar,
    colormap name={jet},
    colorbar style={
    	ylabel={Counts}, y unit={}, x unit={},
    	ymode=log,
    	log ticks with fixed point,
    	ytick={1,2,4,10,20,40,100,200},
    	axis equal=false,
    },
    point meta min=1,
    point meta max=260,
    xlabel={$I$}, x unit={arb.~unit},
    ylabel={$Q$}, y unit={arb.~unit},
    scaled ticks={real:1e3},
    xtick scale label code/.code={},
    ytick scale label code/.code={},
    grid=major,
    axis on top,
    axis equal,
    xmin=-20000,xmax=130000,
    ymin=-90000,ymax=52000,
    ]
    
    \addplot graphics[
    xmin=-89539,xmax=113017,
    ymin=-106428,ymax=51864
    ] {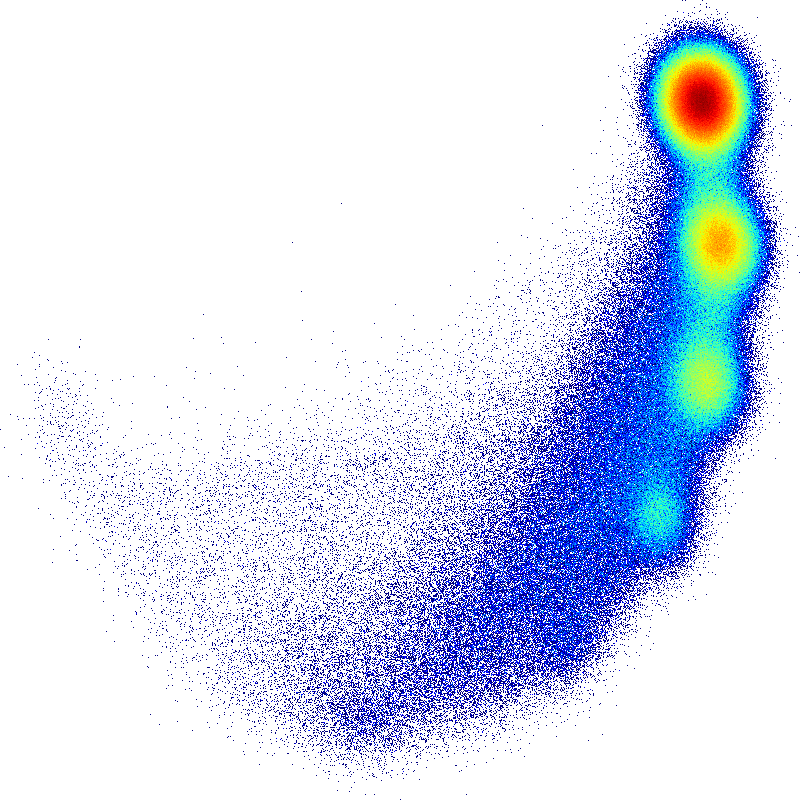};
    
    \node[anchor=south west] at (axis cs:105000,30000) {$\ket{0}$};
    \node[anchor=south west] at (axis cs:105000,0) {$\ket{1}$};
    \node[anchor=north west] at (axis cs:100000,-25000) {$\ket{2}$};
    \node[anchor=north west] at (axis cs:85000,-55000) {$\ket{3}$};
    \end{axis}
    \end{tikzpicture}
    \caption{Histogram showing the in-phase and quadrature response $I$ and $Q$ of one million measurements of a physical Transmon qubit. Please be aware that the color scale is logarithmic. The data is collected by the user task shown in \cref{listing:basic-iqclouds-task}.}
    \label{fig:iqclouds-transmon}
\end{figure}
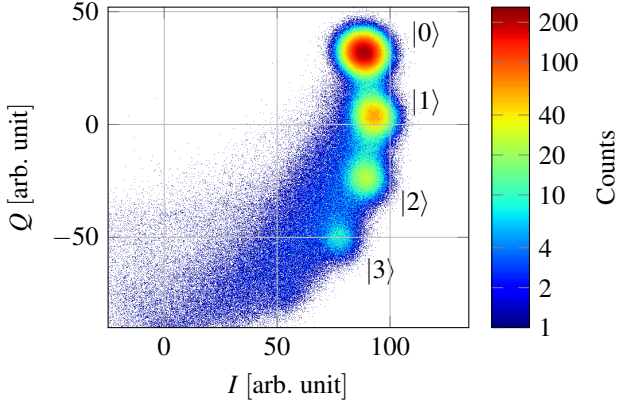
Another important aspect is the necessity to perform further statistics of single measurements.
Due to the latency, the Python client is clearly not suited to perform evaluations based on single measurements.
The Taskrunner, on the other hand, can collect and aggregate single measurement results online and send the collected data independent of the FPGA execution to the user.
This renders experiments like generating a histogram from one million single qubit measurements possible.
The in-phase and quadrature response of the system can be visualized, corresponding to roughly \SI{8}{\mega\byte} of collected data.
An exemplary measurement result from our platform is shown in \cref{fig:iqclouds-transmon}.
Despite the noise, one can clearly see the discrete nature of the single measurement outcome with the probability to end up in different states. 
For this example, data collection on the Taskrunner and subsequent transfer to the Python client took \SI{11,67 \pm 0,06}{\second}. 
The duration is again limited by the intended qubit relaxation delay being \SI{10}{\micro\second} in this experiment. 
With the communication overhead to transfer and start single measurements being roughly \SI{5}{\milli\second}, one can estimate the duration of performing the same experiment on the Python client to be at least $\SI{5000}{\second} \approx \SI{1,4}{\hour}$.

\subsubsection{Calculation of correlation functions}
Also more sophisticated scenarios render possible with the heterogeneous architecture. An example is the calculation of the second order correlation function $g_2$:
\begin{equation}
    g_2 (t) \propto \dfrac{1}{N} \sum_{i=1}^N \int S_{1,i}^\ast (t) S_{1,i}^\ast (t + \tau) S_{2,i}(t + \tau) S_{2,i}(t) \text{d} \tau
\end{equation}
where the signals $S_1$ and $S_2$ of two signal paths are recorded and correlated. As the sum does not commute with the multiplication, single signals need to be processed and multiplied prior to the averaging taking place. This results in a demanding task which is typically implemented on the FPGA.
In order to stay flexible for a variety of different experiments, we perform this calculation and averaging on the Taskrunner. This allows the user to quickly implement and adapt the experiment script to its needs without requiring knowledge in FPGA design. On the downside, the implementation on the real-time processor leads to an increased dead-time between single recordings but is still much faster than if the processing would take place outside of the platform. The implementation on the Taskrunner also allows us to exploit the convolution theorem $\mathcal{F}\{f(t)\ast r(t)\} = F(f)\cdot R(f)$ and employ a fast Fourier transform (FFT) to speed up the calculation and further reduce the measurement dead-time.

As an example, let us assume a $g_2$ correlation measurement with \num{1024} samples for each $S_i$. Each sample consists out of two \SI{16}{\bit} values for the in-phase and quadrature components. The FPGA already reduced the ADC data rate to one value pair every \SI{100}{\nano\second}. This means that a measurement on the FPGA takes $\num{1024} \times \SI{100}{\nano\second} = \SI{102,4}{\micro\second}$. Afterwards, the $g_2$ calculation is performed on the Taskrunner.
To obtain sufficient statistics, we perform \num{100000} averages.
Without the FFT, it takes \SI{1810}{\second} to calculate and average $g_2(t)$ for the first \num{11} values of $t$, i.e. $t \in \{0, \SI{100}{\nano\second}, \ldots, \SI{1000}{\nano\second}\}$.
The same experiment with FFT, in contrast, gives the full $g_2(t)$ with all \SI{1024} possible $t$ values after only \SI{169}{\second}.

Still, the calculation overhead is dominant as the measurement time of \SI{102,4}{\micro\second} per iteration and average is an order of magnitude smaller than the calculation time of \SI{1,31}{\milli\second}. This time is currently dominated by the FFT (\SI{530}{\micro\second}) and the process of copying the signal data from the recording modules (\SI{627}{\micro\second}). 
As the AXI4Lite register read access is a bottleneck here, other data transfer methods like direct memory access (DMA), the possibility of burst reads, and the utilization of a separate AXI interface better suited for the RPU will be addressed in our future work.

\section{Conclusion}
We presented our Taskrunner framework for applications in quantum computing.
It combines a real-time co-processor (RPU) and an application processor (APU) to provide low-latent access to the FPGA logic.
Thereby, we leverage fast parallel processing on the FPGA with the flexibility of a software-based approach.
Due to the usage of the C language, complex functionality and control sequences can be implemented and rapidly tested, with no FPGA design knowledge and testing required.
The integration of the compiler on the APU furthermore leads to less dependencies for the user and enables on-the-fly compilation.

The Taskrunner framework has been characterized by timing benchmarks.
We verified deterministic execution behavior and low-latent AXI register access to the FPGA logic on the order of \SI{300}{\nano\second}.
This is three orders of magnitude faster than if the access to the FPGA logic accessed by an external control computer.
We also motivated different experimental applications that are already utilized in research with superconducting qubits.

Improvements will further reduce the latency, extend the handling of experimental tasks and add automatic calibration features for superconducting qubit experiments.

\section*{Acknowledgment}
Funding was provided by the Initiative and Networking Fund of the Helmholtz Association, within the Helmholtz Future Project ‘Scalable solid state quantum computing’.
Richard Gebauer acknowledges support by the State Graduate Sponsorship Program (LGF) and the Helmholtz International Research School for Teratronics (HIRST).
Nick Karcher acknowledges support by the Karlsruhe School of Elementary Particle and Astroparticle Physics (KSETA).
We are grateful for the experimental support and infrastructure of the Institute of Physics at Karlsruhe Institute of Technology (KIT).
We acknowledge Qkit \cite{Qkit} for providing a convenient measurement software framework for applications in quantum computing with superconducting quantum bits.


\bibliographystyle{IEEEtranS}
\bibliography{bibliography}

\end{document}